\documentclass[copyright]{eptcs}
\providecommand{\event}{T. Neary, D. Woods, A.K. Seda and N. Murphy (Eds.):\\ The Complexity of Simple Programs 2008.}
\providecommand{\volume}{1}
\providecommand{\anno}{2009}
\providecommand{\firstpage}{215}
\usepackage{breakurl}
\usepackage{amsmath}
\usepackage{amsfonts}
\usepackage{amssymb}
\usepackage{algorithmic}
\usepackage{psfrag}
\usepackage{subfig}
\usepackage{wrapfig}

    \usepackage[dvips]{graphicx}  

\newtheorem{theorem}{Theorem}[section]
\newtheorem{lemma}[theorem]{Lemma}

\newenvironment{definition}{{\bf Definition~}}{}

\newtheorem{openproblem}[theorem]{Open Problem}

\newcommand{\Z}{\mathbb{Z}}

\newcommand{\pfunc}[3]{#1 : #2 \dashrightarrow #3 }

\newcommand{\dom}{{\rm dom} \;}

\newcommand{\termasm}[1]{\mathcal{A}_{\Box}\left[\mathcal{#1}\right]}
\newcommand{\prodasm}[1]{\mathcal{A}\left[\mathcal{#1}\right]}
\newcommand{\fgg}[1]{G^\#_{#1}}

\title{Self-Assembly of Infinite Structures}

\author{Matthew J. Patitz \qquad \quad Scott M. Summers\footnote{This author's research was supported in part by NSF-IGERT Training Project in Computational Molecular Biology Grant number DGE-0504304.}
    \email{mpatitz@cs.iastate.edu  \qquad summers@cs.iastate.edu}
    \institute{Department of Computer\\Science, Iowa State University, Ames, IA 50011, USA.\thanks{This research supported in part by National Science Foundation grants 0652569, and 0728806.}}
}
\def\titlerunning{Self-Assembly of Infinite Structures}
\def\authorrunning{M.J. Patitz and S.M. Summers}

\begin{document}

\maketitle
\begin{abstract}
We review some recent results related to the self-assembly of
infinite structures in the Tile Assembly Model.  These results
include impossibility results, as well as novel tile assembly
systems in which shapes and patterns that represent various notions
of computation self-assemble. Several open questions are also
presented and motivated.
\end{abstract}






\section{Introduction}

The simplest mathematical model of nanoscale self-assembly is the
Tile Assembly Model (TAM), an effectivization of Wang tiling
\cite{Wang61,Wang63} that was introduced by Winfree \cite{Winf98}
and refined by Rothemund and Winfree \cite{RotWin00,Roth01}.  (See
also \cite{Adle99,Reif02,SolWin07}.) As a basic model for the
self-assembly of matter, the TAM has allowed researchers to explore
an assortment of avenues into both laboratory-based and theoretical
approaches to designing systems that self-assemble into desired
shapes or autonomously coalesce into patterns that, in doing so,
perform computations.

Actual physical experimentation has driven lines of research
involving kinetic variations of the TAM to deal with molecular
concentrations, reaction rates, etc. as in
\cite{Winfree98simulationsof}, as well as work focused on error
prevention and error correction \cite{ChenGoel04,WinBek03,SolWin05}.
For examples of the impressive progress in the physical realization
of self-assembling systems, see \cite{RoPaWi04,MajSahLaBRei06}.

Divergent from, but supplementary to, the laboratory work, much
theoretical research involving the TAM has also been carried out.
Interesting questions concerning the minimum number of tile types
required to self-assemble shapes have been addressed in
\cite{SolWin07,RotWin00,ACGHKMR02,AGKS04}. Different notions of
running time and bounds thereof were explored in
\cite{AdChGoHu01,BeckerRR06,CGM04}. Variations of the model where
temperature values are intentionally fluctuated and the ensuing
benefits and tradeoffs can be found in \cite{KS07,AGKS04}. Systems
for generating randomized shapes or approximations of target shapes
were investigated in \cite{KaoSchS08,BeckerRR06}.  This is just a
small sampling of the theoretical work in the field of algorithmic
self-assembly.

However, as different as they may be, the above mentioned lines of
research share a common thread. They all tend to focus on the
self-assembly of \emph{finite} structures. Clearly, for
experimental research, this is a necessary limitation. Further, if
the eventual goal of most of the theoretical research is to enable
the development of fully functional, real world self-assembly
systems, a valid question is: ``Why care about anything other than
finite structures?''  This is the question that we address in this
paper.

This paper surveys a collection of recent findings related to the
self-assembly of \emph{infinite} structures in the TAM. As a
theoretical exploration of the TAM, this collection of results seeks
to define absolute limitations on the classes of shapes that
self-assemble. These results also help to explore how fundamental
aspects of the TAM, such as the inability of spatial locations to be
reused and their immutability, affect and limit the constructions
and computations that are achievable.

In addition to providing concise statements and intuitive
descriptions of results, we also define and motivate a set of open
questions in the hope of furthering this line of research. First, we
begin with some preliminary definitions and constructions that will
be referenced throughout this paper.

\section{Preliminaries}

\subsection{The Tile Assembly Model}

This section provides a very brief overview of the TAM. See
\cite{Winf98,RotWin00,Roth01,jSSADST} for other developments of the
model. Our notation is that of \cite{jSSADST}. We work in the
$2$-dimensional discrete space $\Z^2$.   We write $U_2$ for the set
of all {\it unit vectors}, i.e., vectors of length 1 in
$\mathbb{Z}^2$. We write $[X]^2$ for the set of all $2$-element
subsets of a set $X$. All {\it graphs} here are undirected graphs,
i.e., ordered pairs $G = (V, E)$, where $V$ is the set of {\it
vertices} and $E \subseteq [V]^2$ is the set of {\it edges}.

A {\it grid graph} is a graph $G = (V, E)$ in which $V \subseteq \Z^2$ and
every edge $\{\vec{a}, \vec{b} \} \in E$ has the property that $\vec{a} -
\vec{b} \in U_2$.  The {\it full grid graph} on a set $V \subseteq \Z^2$ is the
graph $\fgg{V} = (V, E)$ in which $E$ contains {\it every} $\{\vec{a}, \vec{b}
\} \in [V]^2$ such that $\vec{a} - \vec{b} \in U_2$.

Intuitively, a tile type $t$ is a unit square that can be translated, but not
rotated, having a well-defined ``side $\vec{u}$'' for each $\vec{u} \in U_2$.
Each side $\vec{u}$ of $t$ has a ``glue'' of ``color''
$\textmd{col}_t(\vec{u})$ - a string over some fixed alphabet $\Sigma$ - and
``strength'' $\textmd{str}_t(\vec{u})$ - a natural number - specified by its
type $t$. Two tiles $t$ and $t'$ that are placed at the points $\vec{a}$ and
$\vec{a}+\vec{u}$ respectively, {\it bind} with {\it strength}
$\textmd{str}_t\left(\vec{u}\right)$ if and only if
$\left(\textmd{col}_t\left(\vec{u}\right),\textmd{str}_t\left(\vec{u}\right)\right)
=
\left(\textmd{col}_{t'}\left(-\vec{u}\right),\textmd{str}_{t'}\left(-\vec{u}\right)\right)$.

Given a set $T$ of tile types, an {\it assembly} is a partial function
$\pfunc{\alpha}{\Z^2}{T}$.
An assembly is $\tau$-{\it stable}, where $\tau \in \mathbb{N}$, if it cannot
be broken up into smaller assemblies without breaking bonds whose strengths
sum to at least $\tau$.

Self-assembly begins with a {\it seed assembly} $\sigma$ and
proceeds asynchronously and nondeterministically, with tiles
adsorbing one at a time to the existing assembly in any manner that
preserves stability at all times. A {\it tile assembly system} ({\it
TAS}) is an ordered triple $\mathcal{T} = (T, \sigma, \tau)$, where
$T$ is a finite set of tile types, $\sigma$ is a seed assembly with
finite domain, and $\tau$ is the temperature. An {\it assembly
sequence} in a TAS $\mathcal{T} = (T, \sigma, 1)$ is a (possibly
infinite) sequence $\vec{\alpha} = ( \alpha_i \mid 0 \leq i < k )$
of assemblies in which $\alpha_0 = \sigma$ and each $\alpha_{i+1}$
is obtained from $\alpha_i$ by the ``$\tau$-stable'' addition of a
single tile.
We write $\prodasm{T}$ for the {\it set of all producible assemblies
of} $\mathcal{T}$. An assembly $\alpha$ is {\it terminal}, and we
write $\alpha \in \termasm{\mathcal{T}}$, if no tile can be stably
added to it. We write $\termasm{T}$ for the {\it set of all terminal
assemblies of } $\mathcal{T}$. A TAS ${\mathcal T}$ is {\it
directed}, or {\it produces a unique assembly}, if it has exactly
one terminal assembly i.e., $|\termasm{T}| = 1$. The reader is
cautioned that the term ``directed" has also been used for a
different, more specialized notion in self-assembly \cite{AKKR02}.

A set $X \subseteq \Z^2$ {\it weakly self-assembles} if there exists a TAS
${\mathcal T} = (T, \sigma, 1)$ and a set $B \subseteq T$ such that
$\alpha^{-1}(B) = X$ holds for every assembly $\alpha \in \termasm{T}$.  A set
$X$ {\it strictly self-assembles} if there is a TAS $\mathcal{T}$ for which
every assembly $\alpha\in\termasm{T}$ satisfies $\dom \alpha = X$. The reader
is encouraged to consult \cite{SolWin07} for a detailed discussion of {\it
local determinism} - a general and powerful method for proving the correctness
of tile assembly systems.

\subsection{Discrete Self-Similar Fractals}
In this subsection we introduce discrete self-similar fractals, and
zeta-dimension.

\begin{definition}
\label{def-c-discrete-self-similar-fractal} Let $1 < c \in \mathbb{N}$, and
$X\subsetneq \mathbb{N}^2$. We say that $X$ is a $c$-{\it discrete self-similar
fractal}, if there is a (non-trivial) set $V \subseteq
\{0,\ldots,c-1\}\times\{0,\ldots,c-1\}$ such that $\displaystyle X =
\bigcup_{i=0}^{\infty}{X_i}$, where $X_i$ is the $i^{\textmd{th}}$ {\it stage}
satisfying $X_0 = \{(0,0)\}$, and $X_{i+1} = X_i \cup \left(X_i + c^i V
\right)$. In this case, we say that $V$ {\it generates} $X$.
\end{definition}


\begin{figure}[htp]
\begin{center}
  \subfloat[$X_0$]{\label{fig:fractal_stage1}\includegraphics[width=0.3in]{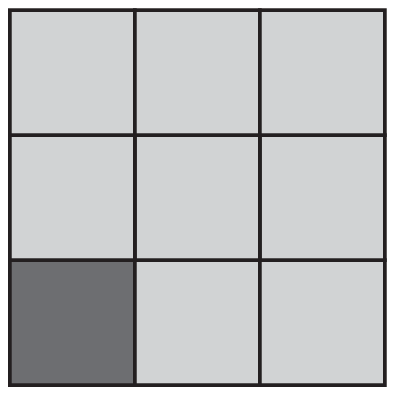}\quad}
  \quad\quad
  \subfloat[$V=X_1$]{\label{fig:fractal_stage2}\includegraphics[width=0.3in]{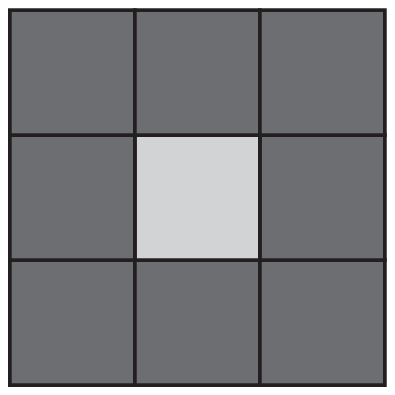}\quad\quad\quad}
  \quad\quad
  \subfloat[$X_2$]{\label{fig:fractal_stage3}\includegraphics[width=0.9in]{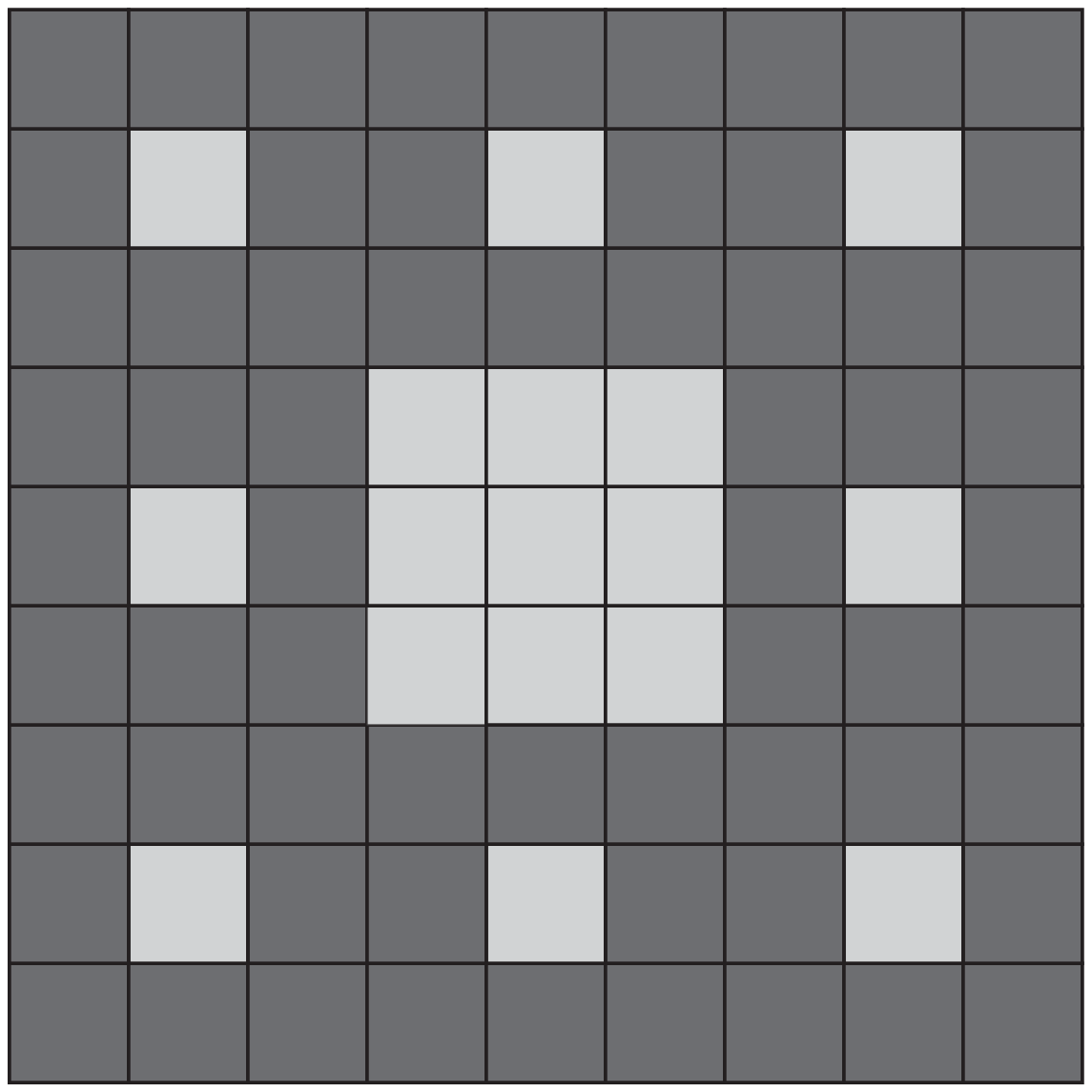}\quad}
  \quad\quad
  \subfloat[$X_3$ (scaled down)]{\label{fig_fractal_stage4}\includegraphics[width=0.9in]{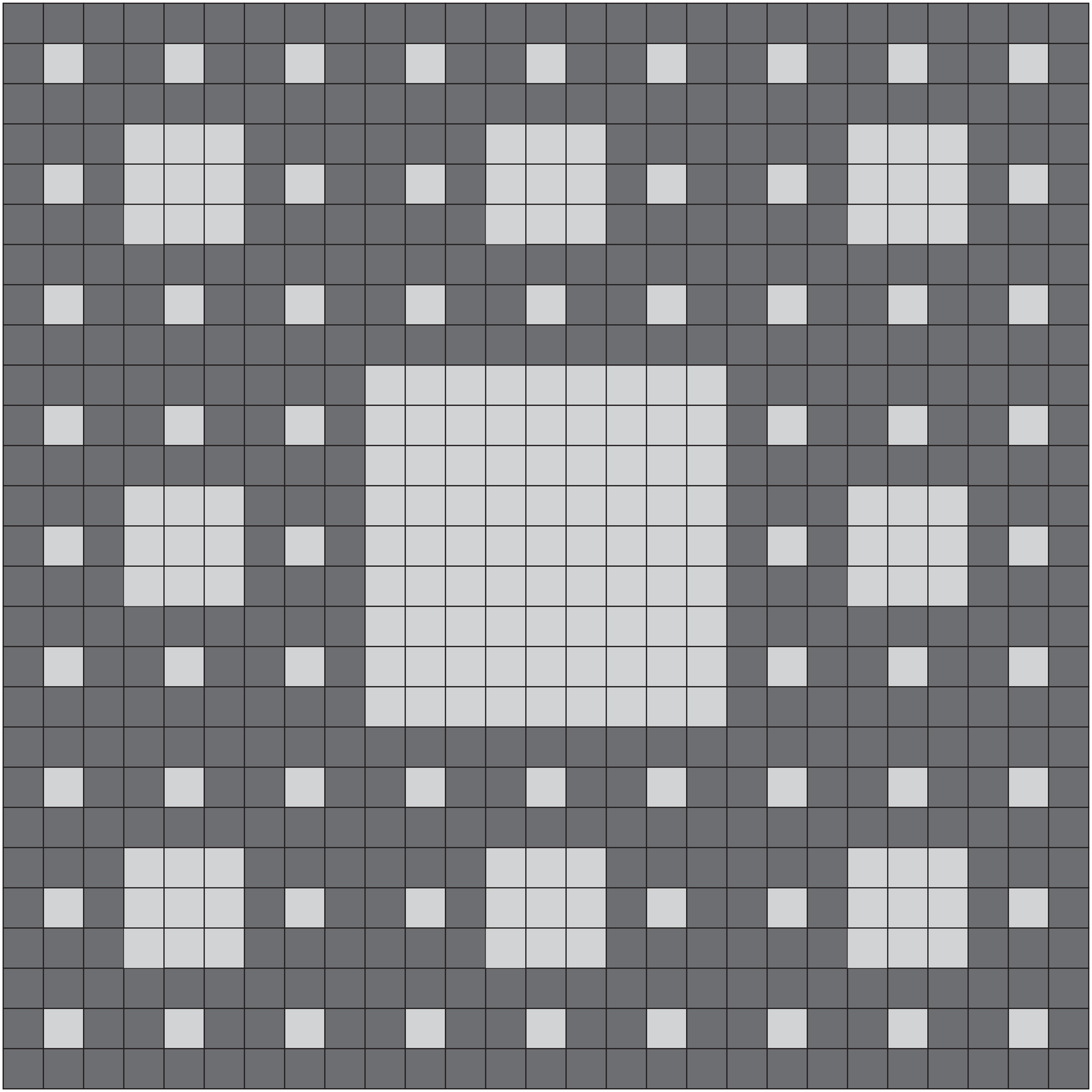}\quad}
  \caption{\small Example of a $c$-discrete self-similar fractal ($c = 3$), the Sierpinski carpet}
  \label{fig:fractals}
\end{center}
\end{figure}

The most commonly used dimension for discrete fractals is zeta-dimension, which
we refer to in this paper.

\begin{definition}\cite{ZD}
For each set $A \subseteq \Z^2$, the {\it zeta-dimension} of $A$ is
\begin{gather*}
     \textmd{Dim}_\zeta(A) = \limsup_{n \rightarrow \infty}\frac{ \log|A_{\le n}|}{\log n},
\end{gather*}
where $A_{\le n} = \{(k,l) \in A \mid  |k|+|l| \le n\}$.
\end{definition}
It is clear that $0 \le \textmd{Dim}_\zeta(A) \le 2$ for all $A
\subseteq \Z^2$.

\subsection{The Wedge Construction}

\begin{wrapfigure}{r}{.5\textwidth}
\includegraphics[width=2.5in]{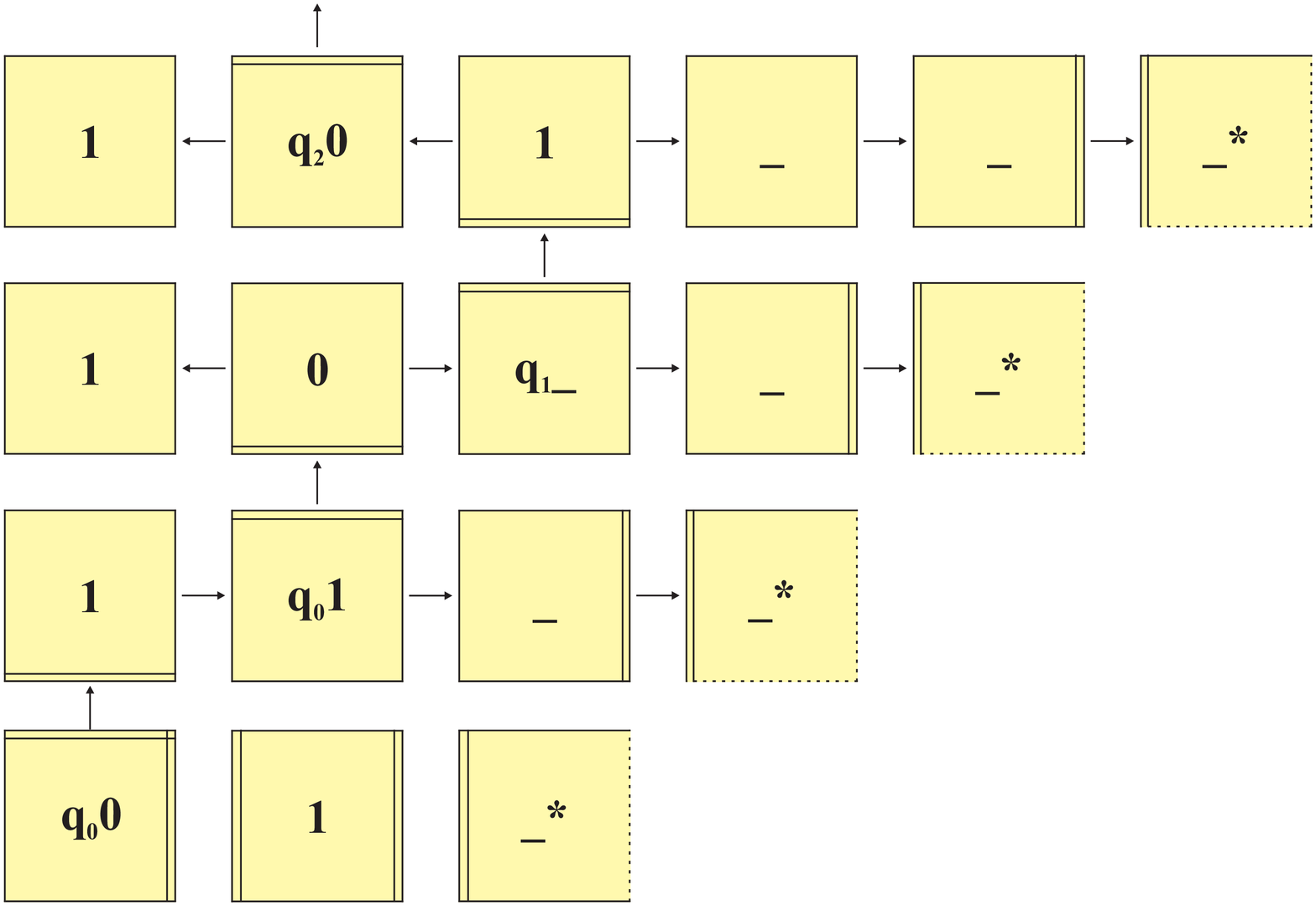}
\caption{Example of the first four rows of a sample wedge construction which is
simulating a Turing machine $M$ on the input string `01'}
\label{fig:wedge_construction}
\end{wrapfigure}

In order to perform universal computation in the TAM, we make use of
a particular TAS called the ``wedge construction'' \cite{SADS}. The
wedge construction, based on Winfree's proof of the universality of
the TAM \cite{Winf98}, is used to simulate an arbitrary Turing
machine $M = (Q, \Sigma, \Gamma, \delta, q_0, q_A, q_R)$ on a given
input string $w \in \Sigma^*$ in a temperature 2 TAS.

The wedge construction works as follows.  Every row of the assembly
specifies the complete configuration of $M$ at some time step.
$M$ starts in its initial state with the tape head on the leftmost
tape cell and we assume that the tape head never moves left off the
left end of the tape. The seed row (bottommost) encodes the initial
configuration of $M$. There is a special tile representing a blank
tape symbol as the rightmost tile in the seed row.  Every subsequent
row grows by one additional cell to the right. This gives the
assembly the wedge shape responsible for its name.
Figure~\ref{fig:wedge_construction} shows the first four rows of a
wedge construction for a particular TM, with arrows depicting a
possible assembly sequence.

\section{Strict Self-Assembly}

The self-assembly of shapes (i.e., subsets of $\mathbb{Z}^2$) in the TAM is
most naturally characterized by strict self-assembly.
In searching for absolute limitations of strict self-assembly in the
TAM, it is necessary to consider infinite shapes because any finite,
connected shape strictly self-assembles via a spanning tree
construction in which there is a unique tile type created for each
point.  In this section we discuss (both positive and negative)
results pertaining to the strict self-assembly of infinite shapes
in the TAM.

\subsection{Pinch-point Discrete Self-Similar Fractals Do Not Strictly Self-Assemble}

In \cite{SADSSF}, Patitz and Summers defined a class $\mathcal{C}$
of (non-tree) ``pinch-point'' discrete self-similar fractals, and
proved that if $X \in \mathcal{C}$, then $X$ does not strictly
self-assemble.

%
%
%

\begin{definition}
A {\it pinch-point discrete self-similar fractal} is a discrete self-similar
fractal satisfying (1) $\{(0,0),(0,c-1),(c-1,0)\} \subseteq V$, (2) $V \cap
(\{1,\ldots c-1\}\times \{c-1\}) = \emptyset$, (3), $V \cap (\{c-1\} \times
\{1,\ldots, c-1\}) = \emptyset$, and $\fgg{V}$ is connected
\end{definition}

\begin{wrapfigure}{r}{.45\textwidth}
    \begin{center}
    \includegraphics[width=2.0in]{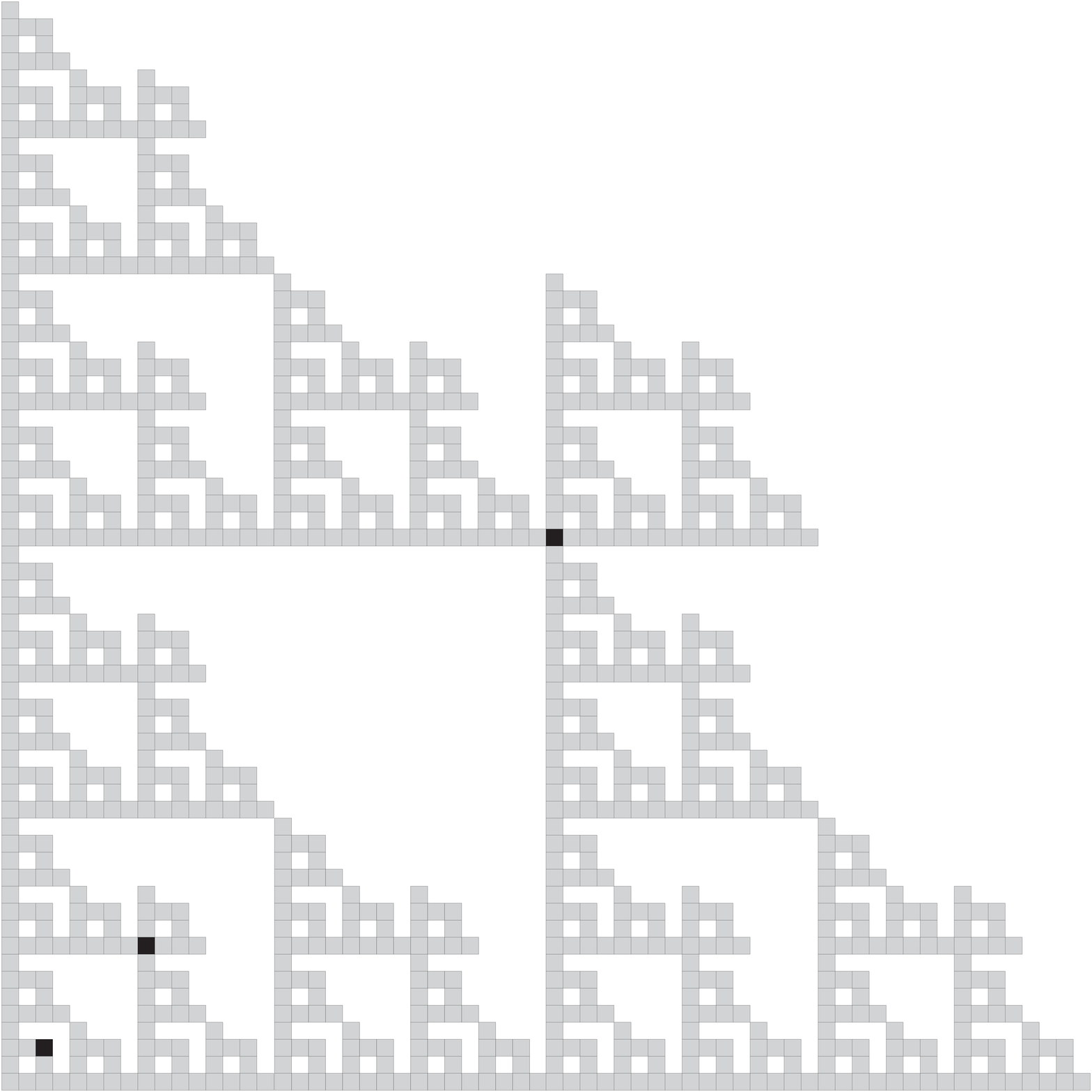}
    \caption{\label{fig:pinch_points_highlighted}\small An example of the first four stages of pinch-point fractal with the first three
    pinch-points highlighted in black.}
    \end{center}
\end{wrapfigure}

A famous example of a pinch-point fractal is the standard discrete
Sierpinski triangle $\mathbf{S}$. The impossibility of the strict
self-assembly of $\mathbf{S}$ was first shown in \cite{jSSADST}.
Figure~\ref{fig:pinch_points_highlighted} shows another example of a
pinch-point discrete self-similar fractal. Note that any fractal $X$
such that $\fgg{X}$ is a tree is necessarily a pinch-point discrete
self-similar fractal.


The following (slight) generalization to \cite{jSSADST} was shown in
\cite{SADSSF}.

\begin{theorem}
\label{firstmaintheorem} If $X \subsetneq \mathbb{N}^2$ is a pinch-point
discrete self-similar fractal, then $X$ does not strictly self-assemble in the
TAM.
\end{theorem}

The idea behind the proof of Theorem~\ref{firstmaintheorem} can be seen in
Figure~\ref{fig:pinch_points_highlighted}. Note that the black points are
pinch-points in the sense that arbitrarily large aperidic sub-structures appear
on the far-side of the black tile from the origin.


Theorem~\ref{firstmaintheorem} motivates the following question.

\begin{openproblem}\label{no_fractals_assemble}
Does any non-trivial discrete self-similar fractal strictly self-assemble in
the TAM?  We conjecture that the answer is `no', for any temperature $\tau \in
\mathbb{N}$.
However, proving that there exists a (non-trivial) discrete self-similar
fractal that does strictly self-assemble would likely involve a novel and
useful algorithm for directing the behavior self-assembly.
\end{openproblem}

\subsection{Strict Self-Assembly of Nice Discrete Self-Similar Fractals}

As shown above, there is a class of discrete self-similar fractals that do not
strictly self-assemble (at any temperature) in the TAM.  However, in
\cite{SADSSF}, Patitz and Summers introduced a particular set of ``nice''
discrete self-similar fractals that contains some but not all pinch-point
discrete self-similar fractals. Further, they proved that any element of the
former class has a ``fibered'' version that strictly self-assembles.

\subsubsection{Nice Discrete Self-Similar Fractals}

\begin{definition}
A {\it nice discrete self-similar fractal} is a discrete
self-similar fractal such that $(\{0,\ldots,c-1\} \times \{0\}) \cup
(\{0\}\times\{0,\ldots,c-1\}) \subseteq V$, and $\fgg{V}$ is
connected.
\end{definition}

\begin{figure}[h]
  \begin{center}
  \subfloat[Nice]{\label{fig:nice}\includegraphics[width=1.25in]{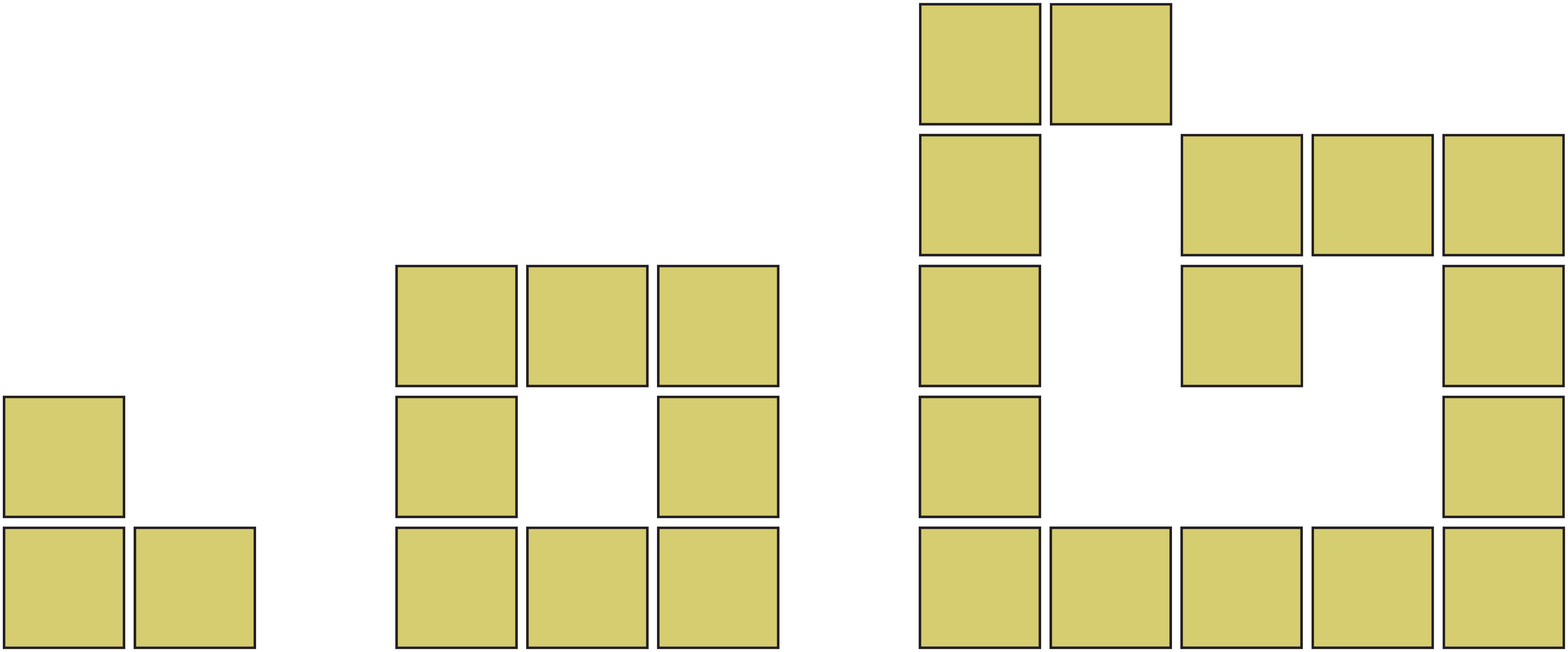}}
  \quad\quad
  \subfloat[Non-nice]{\label{fig:bad}\includegraphics[width=0.73in]{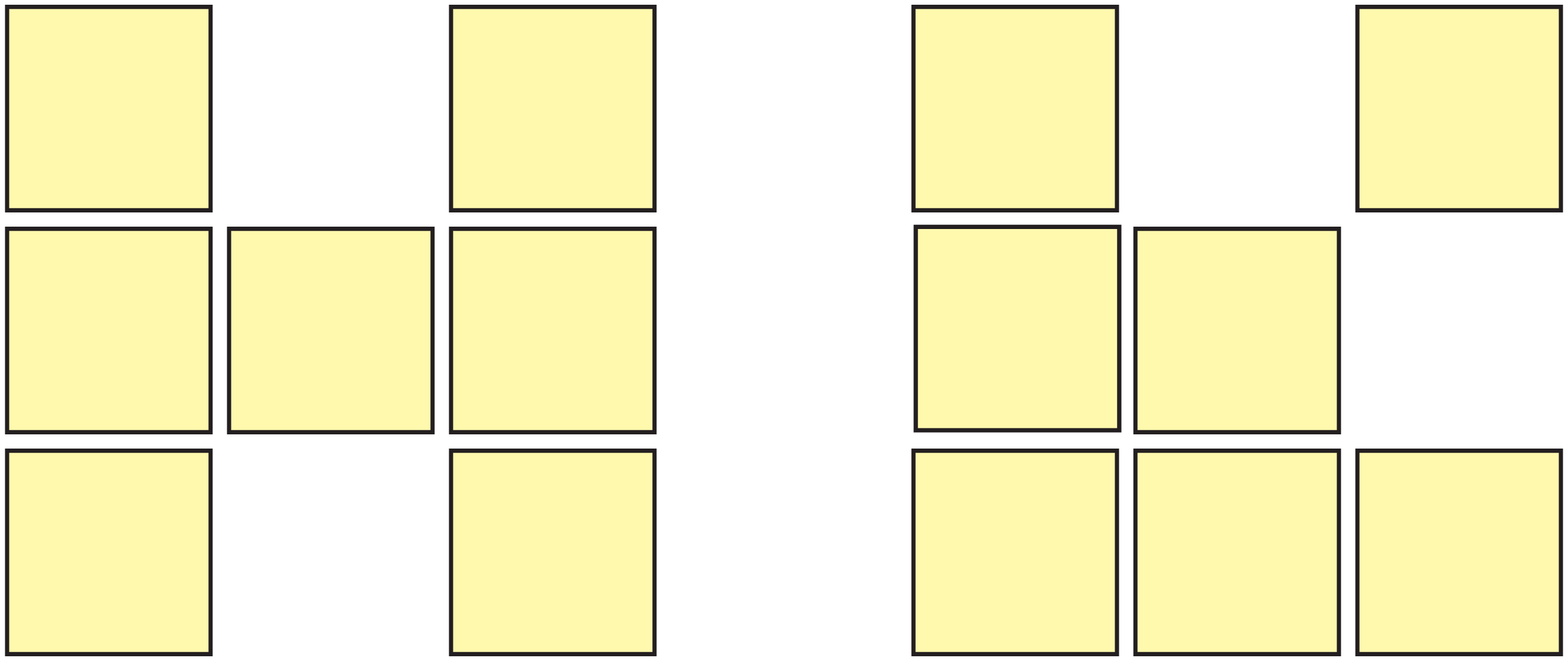}}
  \caption{\small Stage 2 of some discrete self-similar fractals.}
  \label{fig:nice_fractals}
  \vspace{-20pt}
  \end{center}
\end{figure}

See Figure~\ref{fig:nice_fractals} for examples of both nice, and non-nice
discrete self-similar fractals.

\subsubsection{Nice Fractals Have Fibered Versions}

The inability of pinch-point fractals (and the conjectured inability of any
discrete self-similar fractal) to strictly self-assemble in the TAM is based on
the intuition that the necessary amount of information cannot be transmitted
through available connecting tiles during self-assembly.

Thus, for any nice discrete self-similar fractal $X$, Patitz and Summers
\cite{SADSSF} defined a fibered operator $\mathcal{F}(X)$ (a routine extension
of \cite{jSSADST}) which adds, in a zeta-dimension-preserving manner,
additional bandwidth to $X$. Strict self-assembly of $\mathcal{F}(X)$ is
achieved via a ``modified binary counter'' algorithm that is embedded into the
additional bandwidth of $\mathcal{F}(X)$.

\begin{figure}
    \begin{center}
    \includegraphics[width=5.0in]{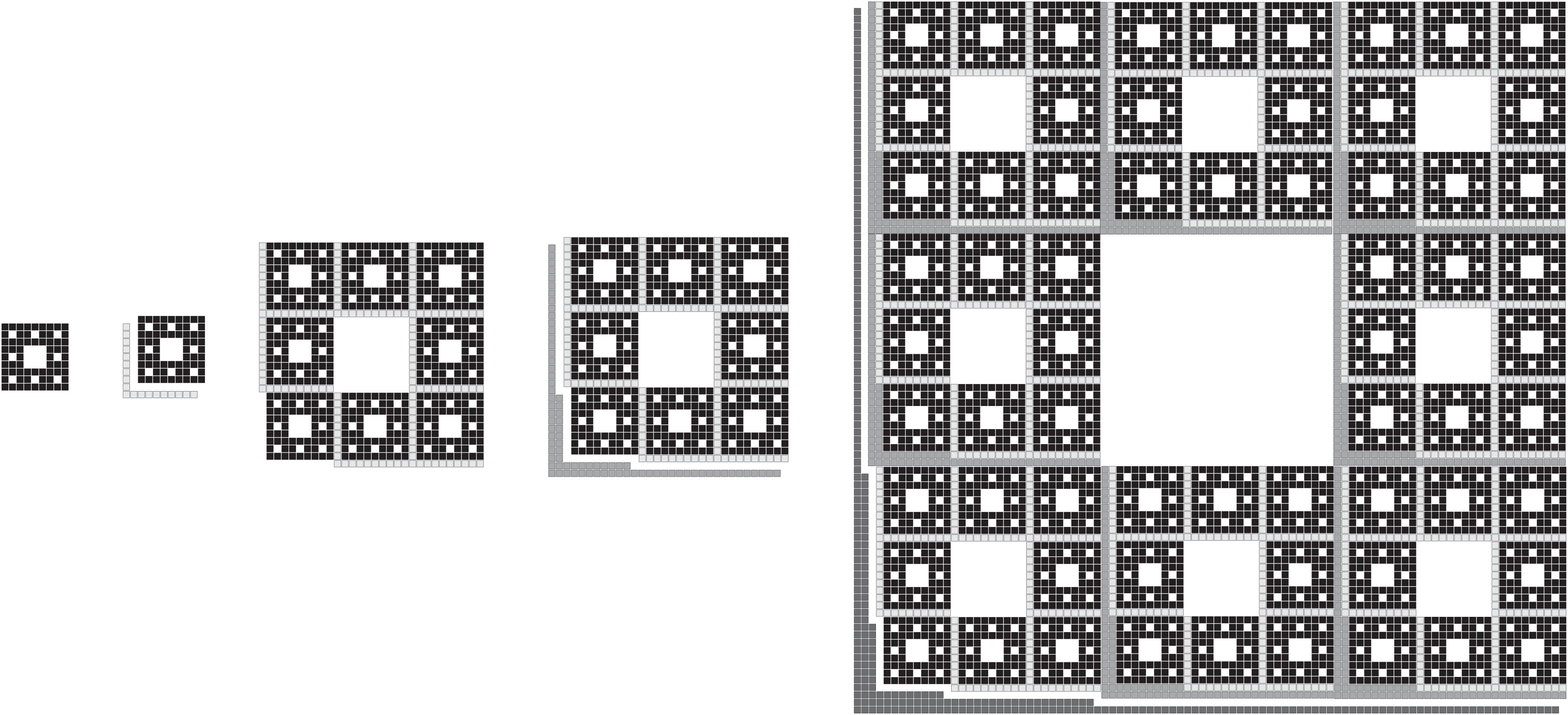}
    \caption{\small Construction of the fibered Sierpinski carpet}
    \label{fig:fibered_example}
    \end{center}
\end{figure}

For any nice discrete self-similar fractal $X$, $\mathcal{F}(X)$ is
defined recursively. Figure~\ref{fig:fibered_example} shows an
example of the construction of $\mathcal{F}(X)$, where $X$ is the
discrete Sierpinski carpet.
Note that $\mathcal{F}(X)$ is only defined if $X$ is a nice discrete
self-similar fractal. Moreover, it appears non-trivial to extend
$\mathcal{F}$ to other discrete self-similar fractals such as the
`H' fractal (the second-to-the-right most image in
Figure~\ref{fig:nice_fractals}).




\begin{openproblem}
Does there exist a zeta-dimension-preserving fibered operator
$\mathcal{F}$ for a class of discrete self-similar fractals
which is a superset of the nice discrete self-similar fractals (e.g.
it also includes the `H' fractal)?  The above open question is
intentionally vague. Not only should $\mathcal{F}$ preserve
zeta-dimension, but $\mathcal{F}(X)$ should also ``look'' like $X$
in some reasonable visual sense.
\end{openproblem}


\section{Weak Self-Assembly}


Weak self-assembly is a natural way to define what it means for a
tile assembly system to compute. There are examples of (decidable)
sets that weakly self-assemble but do not strictly self-assemble
(i.e., the discrete Sierpinski triangle \cite{jSSADST}). However, if
a set $X$ weakly self-assembles, then $X$ is necessarily computably
enumerable. In this section, we discuss results that pertain to the
weak self-assembly of (1) discrete self-similar fractals
\cite{SADSSF}, (2) decidable sets \cite{SADS}, and (3) computably
enumerable sets \cite{CCSA}.

\subsection{Discrete Self-Similar Fractals}

Recall that pinch-point discrete self-similar fractals do not strictly
self-assemble (at any temperature). Furthermore, Patitz and Summers
\cite{SADSSF} proved that \emph{no} (non-trivial) discrete self-similar fractal
weakly self-assembles in a locally deterministic \cite{SolWin07} temperature 1
tile assembly system.



\begin{theorem}
\label{weaktheorem} If $X \subsetneq \mathbb{N}^2$ is a discrete
self-similar fractal, and $X$ weakly self-assembles in the locally
deterministic TAS $\mathcal{T}_X = (T,\sigma,\tau)$, where $\sigma$
consists of a single tile placed at the origin, then $\tau > 1$.
\end{theorem}

Intuitively, the proof relies on the aperiodic nature of discrete
self-similar fractals and the fact that the binding (a.k.a.
adjacency) graph of the terminal assembly of $\mathcal{T}_X$ is an
infinite tree, and every infinite branch is composed of an infinite,
periodically repeating sequence of tile types.

\begin{openproblem}
Does Theorem \ref{weaktheorem} hold for any directed (not necessarily locally
deterministic) TAS?  We conjecture that it does, and that such a proof would
provide useful new tools for impossibility proofs in the TAM.
\end{openproblem}

\subsection{Decidable Sets}

We now shift gears and discuss the weak self-assembly of sets at
temperature~2.

\subsubsection{A Characterization of Decidable Sets of Natural Numbers}
\label{decidable_in_N}

In \cite{SADS}, Patitz and Summers exhibited a novel
characterization of decidable sets of positive integers in terms of
weak self-assembly in the TAM, where they proved the following
theorem.

\begin{theorem}
\label{sads_theorem} Let $A \subseteq \mathbb{N}$. Then $A \subseteq
\mathbb{N}$ is decidable if and only if $A \times \{0\}$ and $A^c
\times \{0\}$ weakly self-assemble.
\end{theorem}

Theorem~\ref{sads_theorem} is the ``self-assembly version'' of the
classical theorem, which says that a set $A \subseteq \mathbb{N}$ is
decidable if and only if $A$ and $A^c$ are computably enumerable.
The following lemma makes the proof of the reverse direction of
Theorem~\ref{sads_theorem} straight-forward.

\begin{lemma} \label{primitive_simulator}
Let $X \subseteq \mathbb{Z}^2$. If $X$ weakly self-assembles, then $X$ is
computably enumerable.
\end{lemma}

The proof of Lemma~\ref{primitive_simulator} constructs a
self-assembly simulator to enumerate $X$.

\begin{figure}[htb]
\psfrag{0}{$M(0)$} \psfrag{1}{$M(1)$} \psfrag{2}{$M(2)$}
\begin{center}
\includegraphics[width=2.5in]{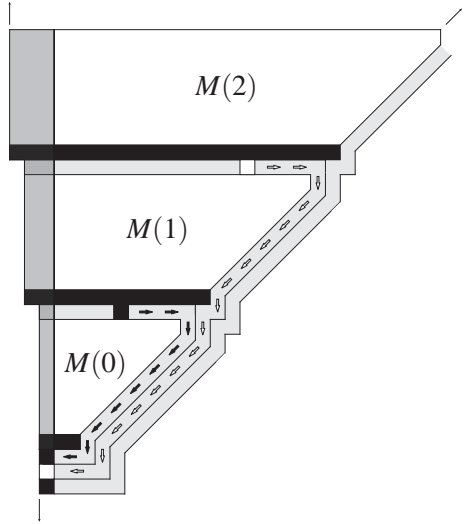}
\end{center}
\caption{\small The left-most (dark grey) vertical bars represent a binary
counter that is embedded into the tile types of the TM; the darkest
(black) rows represent the initial configuration of $M$ on inputs 0,
1, and 2; and the (light grey) horizontal rows that contain a white/black
tile represent halting configurations of $M$. Although this image
seems to imply that the embedded binary counter increases its width
(to the left) each time it increments, this is not true in the
construction. This image merely depicts the general shape of the
counter as it increments.}
\label{fig:decider_overview}
\end{figure}

To prove the forward direction of Theorem~\ref{sads_theorem}, it
suffices to construct an infinite stack of wedge constructions and
simply propagate the halting signals down to the negative $y$-axis.
This is illustrated in Figure~\ref{fig:decider_overview}.

%
%
%

\subsubsection{Quadrant Optimality}

In addition to their positive result, Patitz and Summers \cite{SADS}
established that any tile assembly system $\mathcal{T}$ that
``row-computes'' a decidable language $A \subseteq \mathbb{N}$
having sufficient space complexity must place at least one tile in
each of two adjacent quadrants. A TAS $\mathcal{T}$ is said to
\emph{row-compute} a language $A \subseteq \mathbb{N}$ if
$\mathcal{T}$ simulates a TM $M$ with $L(M) = A$ on every input $n
\in \mathbb{N}$, one row at a time, and uses single-tile-wide paths
of tiles to propagate the answer to the question, ``does $M$ accept
input $n$?" to the $x$-axis. Figure~\ref{fig:decider_overview}
depicts the essence of what it means for a TAS to row-compute some
language. This result, stated precisely, is as follows.

\begin{theorem}\label{quadrant_optimality}
Let $A \subseteq \mathbb{N}$. If $A \not \in
\textmd{DSPACE}\left(2^n\right)$, and $\mathcal{T}$ is any TAS that
``row-computes'' $A$, then every terminal assembly of $\mathcal{T}$
places at least one tile in each of two adjacent quadrants.
\end{theorem}

\begin{openproblem}
Let $A \subseteq \mathbb{N}$ with $A \not \in \textmd{DSPACE}\left(2^n\right)$.
Is it possible to construct a directed TAS $\mathcal{T}$ in which the sets $A
\times \{0\}$ and $A^c \times \{0\}$ weakly self-assemble, and every terminal
assembly $\alpha \in \termasm{T}$ is contained in the first quadrant? We
conjecture that the answer is `no', and any proof would account for all,
possibly exotic methods of computation in the TAM, not only by row-computing.
\end{openproblem}


\subsubsection{There Exists a Decidable Set That Does Not Weakly Self-Assemble}

In contrast to Theorem~\ref{sads_theorem}, Lathrop, Lutz, Patitz, and Summers
\cite{CCSA} proved that there are decidable sets $D \subseteq \mathbb{Z}^2$
that do not weakly self-assemble. To see this, for each $r \in \mathbb{N}$,
define
$$
D_r = \{\left.(m,n) \in \mathbb{Z}^2 \; \right| \; |m|+|n|=r\}.
$$
This set is a ``diamond'' in $\mathbb{Z}^2$ with radius $r$ and
center at the origin. For each $A \subseteq \mathbb{N}$, let
$$
D_A = \bigcup_{r\in A}{D_r}.
$$
This set is the ``system of concentric diamonds'' centered at the
origin with radii in $A$.
Using Lemma~\ref{primitive_simulator}, one can establish the
following result.

\begin{lemma}\label{dtime}
Let $A \in \mathbb{N}$.  If $D_A$ weakly self-assembles, then
there exists an algorithm that, given $r \in \mathbb{N}$, halts
and accepts in time $O(2^{4n})$, where $n = \lfloor\lg r \rfloor + 1$,
if and only if $r \in A$.
\end{lemma}

The proof of Lemma~\ref{dtime} is based on the simple observation
that each diamond is finite, and once a tile is placed at some
point, it cannot be removed.
The time hierarchy theorem \cite{HarSte65} can be employed to show
that there exists a set $A \in \mathbb{N}$ such that $A \in
\textmd{DTIME}\left(2^{5n}\right)-\textmd{DTIME}\left(2^{4n}\right)$.
Lemma~\ref{dtime} with $D=D_A$ is sufficient to prove the following
theorem.

\begin{theorem}\label{CCSA_impossibility_proof}
There is a decidable set $D \subseteq
\mathbb{Z}^2$ that does not weakly self-assemble.
\end{theorem}

It is easy to see that if $A \subseteq \mathbb{N}$, then $D_A \in
\textmd{DTIME}\left(2^{\textmd{linear}}\right)$ because you can
simulate self-assembly with a Turing machine. Is it possible to
do better?

\begin{openproblem} \cite{CCSA}
Is there a polynomial-time decidable set $D \in \mathbb{Z}^2$ such
that $D$ does not weakly self-assemble?
\end{openproblem}


\subsection{Computably Enumerable Sets}

The characterization of decidable sets in terms of weak
self-assembly \cite{SADS} is closely related to the characterization
of computably enumerable sets in terms of weak self-assembly due to
Lathrop, Lutz, Patitz and Summers \cite{CCSA}.

Let $f: \mathbb{Z}^+ \rightarrow \mathbb{Z}^+$ be a function such
that for all $n \in \mathbb{N}$, $f(n) \geq n$ and $f(n) =
O\left(n^2\right)$.
For each set $A\subseteq \mathbb{Z}^+$,
the set
$$
X_A = \left\{ (f(n),0) \mid n \in A \right\}
$$
is thus a straightforward representation of $A$ as a set of points
on the positive $x$-axis. The first main result of \cite{CCSA} is
stated as follows.

\begin{theorem}
\label{CCSA_firstmaintheorem} If $f: \mathbb{Z}^+ \rightarrow
\mathbb{Z}^+$ is a function as defined above, then, for all $A
\subseteq \Z^+$, $A$ is computably enumerable if and only if the set
$X_A = \{ (f(n), 0) \mid n \in A \}$ self-assembles.
\end{theorem}

\begin{figure}[h]
\begin{center}
\includegraphics[width=3.5in]{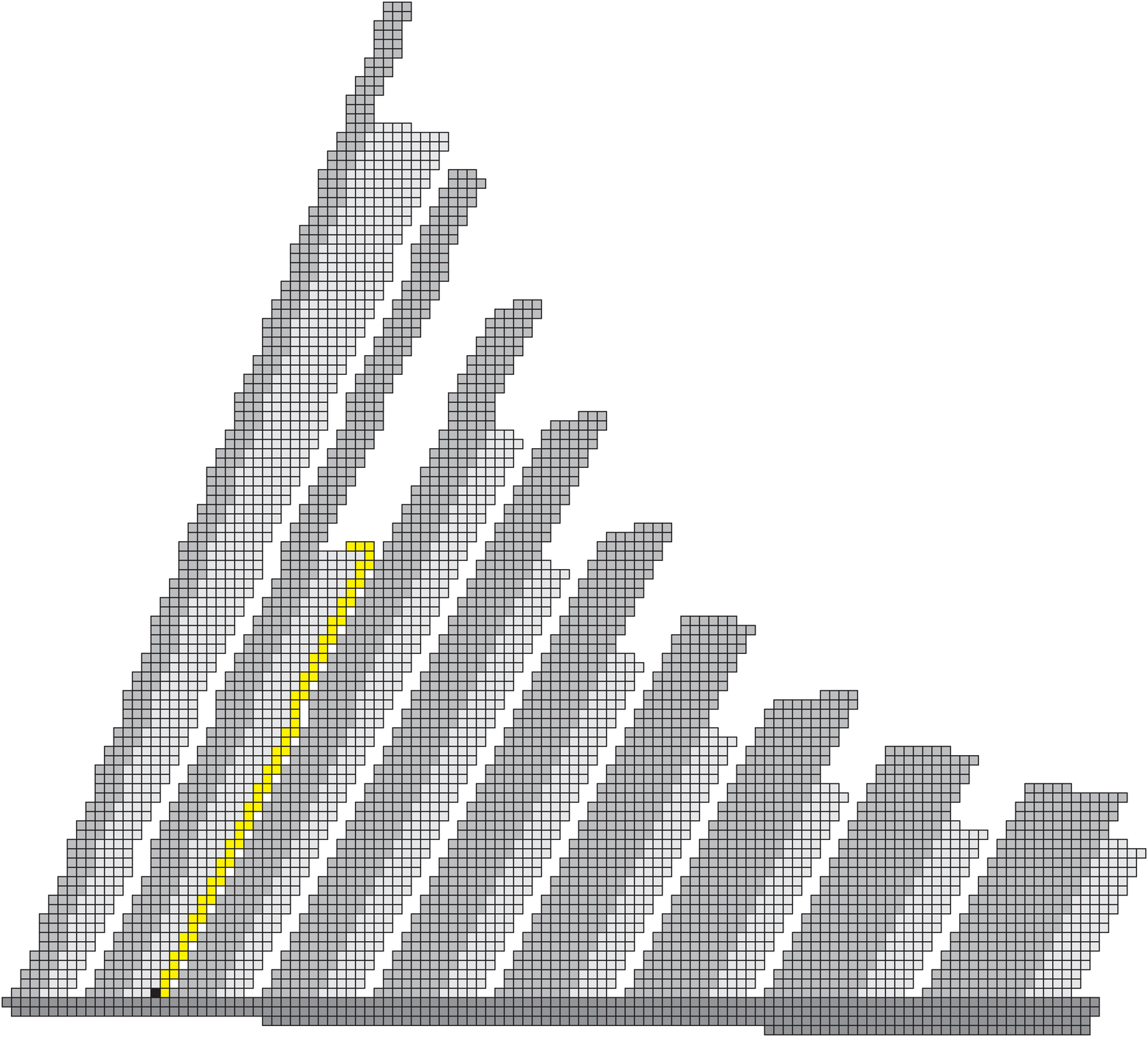}
\caption{Simulation of $M$ on every input $x \in \mathbb{N}$. Notice that
$M(2)$ halts - indicated by the black tile along the $x$-axis.}
\label{fig:CCSA_construction}
\end{center}
\end{figure}

The reverse direction of the proof follows easily from
Lemma~\ref{primitive_simulator}. To prove the forward direction, it is
sufficient to exhibit, for any TM $M$, a directed TAS $\mathcal{T}_M$ that
correctly simulates $M$ on all inputs $x \in \mathbb{Z}^+$ in $\mathbb{Z}^2$. A
snapshot of the main construction of \cite{CCSA} is shown in
Figure~\ref{fig:CCSA_construction}.

Intuitively, $\mathcal{T}_M$ self-assembles a ``gradually thickening
bar'' immediately below the positive $x$-axis with upward growths
emanating from well-defined intervals of points. For each $x \in
\mathbb{Z}^+$, there is an upward growth, in which a modifed wedge
construction carries out a simulation of $M$ on $x$. If $M$ halts on
$x$, then (a portion of) the upward growth associated with the
simulation of $M(x)$ eventually stops, and sends a signal down along
the right side of the upward growth via a one-tile-wide-path of
tiles to the point $(f(x),0)$, where a black tile is placed.

Note that Theorem~\ref{sads_theorem} is exactly
Theorem~\ref{CCSA_firstmaintheorem} with ``computably enumerable'' replaced
with ``decidable,'' and $f(n) = n$.

\begin{openproblem} \cite{CCSA}
Does Theorem~\ref{CCSA_firstmaintheorem} hold for any $f$ such that
$f(n) = O(n)$? We conjecture that the answer is ``no'', and that the
construction of \cite{CCSA} is effectively optimal.  If the answer
to this question is ``yes,'' then the proof would require a novel
construction which manages to provide an infinite amount of space
for each of an infinite number of perhaps non-halting computations
in a more compact way than \cite{CCSA}.
\end{openproblem}


\section{Conclusion}

This paper surveyed a subset of recent theoretical results in algorithmic
self-assembly relating to the self-assembly of infinite structures in the TAM.
Specifically, in this paper we reviewed impossibility results with respect to
the strict/weak self-assembly of various classes of discrete self-similar
fractals \cite{SADSSF}, impossibility results for the weak self-assembly of
exponential-time decidable sets \cite{CCSA}, characterizations of particular
classes of languages in terms of weak self-assembly \cite{CCSA,SADS}, and the
strict self-assembly of fractal-like structures. Finally, we believe that the
benefit of continued research along these lines has the potential to shed light
on the elusive relationship between geometry and computation.

%

\bibliographystyle{eptcs}
\providecommand{\bysame}{\leavevmode\hbox to3em{\hrulefill}\thinspace}
\providecommand{\MR}{\relax\ifhmode\unskip\space\fi MR }
\providecommand{\MRhref}[2]{%
  \href{http://www.ams.org/mathscinet-getitem?mr=#1}{#2}
}
\providecommand{\href}[2]{#2}

\end{document}